\hsize=6.11in \vsize=8.2in
\font\fourteenbold=cmbx14
\font\tenpoint=cmr10
\fourteenbold
\centerline{ Orthodox Quantum Mechanics Free from Paradoxes}
 
\vskip 20pt
\tenpoint
\centerline{Rodrigo Medina}
                       
\centerline{Instituto Venezolano de Investigaciones Cient\'\i ficas IVIC}
\centerline{Apartado 21827, Caracas 1020A, Venezuela}

\vskip 40pt            
\centerline{\bf ABSTRACT}
\vskip 20pt
                            
{\it     A  formulation of quantum mechanics based on an  operational 
definition  of  state  is  presented.   This  formulation,  which 
includes   explicitly  the  macroscopic  systems,   assumes   the 
probabilistic  interpretation and is nevertheless objective. 
The  classical  paradoxes of quantum mechanics are  analyzed  and 
their  origin is found to be the fictitious properties  that  are usually 
attributed to quantum-mechanical states.  The hypothesis that any
mixed state can always be considered as an incoherent superposition
of pure states is found to contradict quantum mechanics. A solution of
EPR paradox
is proposed. It is shown that entanglement of quantum states
is compatible with realism and
locality of events, but implies non-local encoding of 
information.
}
\vskip 40pt
\fourteenbold
\centerline{ M\'ecanique Quantique Orthodoxe Libre de Paradoxes}
\tenpoint
\vskip 20pt
\centerline{\bf R\'ESUM\'E}
\vskip 20pt

{\it
   Nous pr\'esentons une formulation de la m\'ecanique quantique bas\'ee
sur une d\'efini\-tion op\'era\-tion\-nelle d'\'etat. Cette formulation,
qui comprend explicitement les syst\`emes ma\-croscopiques, se base sur
l'interpr\'etation probabiliste et est n\'eanmoins objective. Les
paradoxes classiques de la m\'ecanique quantique sont analys\'es et
leur origine est li\'ee aux propri\'et\'es fictives
qui sont habituellement attribu\'ees aux \'etats quantiques. 
L'hypoth\`ese que n'importe quel \'etat m\'elang\'e peut \^etre
consid\'er\'e toujours comme une superposition incoh\'erente d'\'etats
purs, serait contradictoire. Une solution du paradoxe de EPR est
propos\'ee. Nous montrons que
l'enchev\^etrement des \'etats quantiques est compatible avec le
r\'ealisme et la localit\'e des \'ev\'enements, mais qu'il implique
la codification non-locale de l'information.
}

\vfill\eject

\beginsection I. Introduction  
  
     From  the very beginning of quantum mechanics a  controversy 
about its interpretation developed. Very soon different paradoxes 
and  problems  of  interpretation  appeared, {\it e.g.}~Schr\"odinger's 
cat,$^{(1)}$  the  reduction of the wave-packet,$^{(2)}$  the problem posed 
by Einstein,  Podolsky and Rosen (EPR),$^{(3)}$  etc..  Although most 
physicists  do not question the  Copenhagen 
interpretation,  no general agreement exits with  the 
solutions  that have been proposed for these paradoxes as 
revealed by a large amount of articles on  this subject that continue
to be published. 
  Extensive reviews of related topics are given  by 
Selleri  and  Tarozzi,$^{(4)}$ de Muynck,$^{(5)}$ Cramer,$^{(6)}$
Stapp$^{(7)}$
and Home and Whitaker.$^{(8)}$ To disentangle the issues of
this seventy years old controversy is not a task as simple as
one may believe.
For example, a very diffuse opinion$^{(8,9)}$  defines a whole class of
interpretations of quantum mechanics (ensemble interpretations are called by
Home and Whitaker) as those that accept the following statement:
\item{($I1$)} ``The quantum-mechanical state represents an ensemble of 
similarly prepared systems''.

On the contrary,
we believe that such statement should not be considered an issue at all.
As can be easily verified with an operational analysis, anybody that uses
the standard quantum-mechanical  formalism for computing probabilities
assumes, explicitly or implicitly, $I1$ to be true. This fact is not related
to the interpretation of quantum mechanics, but to the {\bf empirical}
meaning of probability itself.  For us, ensemble interpretation 
and probability interpretation are synonymous.

A question which, instead, we consider a real issue is the origin of
probability
of quantum phenomena. The Orthodox Interpretation assumes that:
\item{($I2^{\prime}$)} ``In quantum mechanics probability is intrinsic.''

On the other side for the Statistical Interpretation:
\item{($I2^{\prime\prime}$)}``Probability in quantum mechanics just
reflects the existence of an underlying reality which is not completely
described by the quantum-mechanical state.''

In other words the statistical interpretation postulates the existence
of a hidden state that represents the physical reality of the single
system. The quantum-mechanical state represents then an ensemble of such
hidden states. Note that this last ensemble is different from the one that
appears in $I1$.  Today it is quite clear that no statistical
interpretation is possible (see for example Mermin$^{(10)}$),
unless one accepts that probability distributions depend on the
measuring apparatus.

Another issue is the completeness of quantum mechanics.
Many sustain that orthodoxy requires accepting that:
\item{($I3^{\prime}$)}``The pure quantum state describes completely
the physical reality of a single system.''

Of course, the statistical interpretation maintains the opposite:
\item{($I3^{\prime\prime}$)}``The quantum state does not describe
completely the physical reality.''

The famous EPR paper was intented to prove $I3^{\prime\prime}$.
As this statement is an obvious consequence of $I2^{\prime\prime}$,
that paper was considered an argument in favor of the statistical
interpretation.
 Here comes a crucial
 point. $I3^{\prime\prime}$ follows from $I2^{\prime\prime}$, but
$I3^{\prime\prime}$ does not imply $I2^{\prime\prime}$. That is, from
the fact that the quantum state does not represent completely the
physical reality does not follow that there is some other kind of
state that does the job. This logical relationship among the statements
has the following consequences: 1) The no-go theorems that invalidate
$I2^{\prime\prime}$ do not invalidate $I3^{\prime\prime}$; 2) the
orthodox interpretation is not forced to assume that $I3^{\prime}$ is
true. Ballentine$^{(9)}$  has shown that the superfluous assumption
$I3^{\prime}$ is the origin of many absurdities that are attributed
to the orthodox interpretation, but, unlike us, he considers instead
valid $I2^{\prime\prime}$ which is also superfluous.

 The orthodox interpretation of quantum mechanics 
has  been labeled as subjectivistic because of the essential  role 
played  by the Observer within the theory. Actually this view is
shared only by some partisans of the Copenhagen school like
von Neumann, others like Bohr and Heisenberg stress the role
of the measuring apparatus.  Such abandonment  of   
realism, {\it i.e.}~the doctrine that the objects that make up the
physical world have an existence and behavior independent of human
mind, was attributed to the operationalist philosophy that  was 
professed  by  the Copenhagen school.$^{(11)}$  In our opinion  the
operational definition of physical concepts is  not in conflict with
realism; instead, by defining
 operationally such fundamental concepts as preparation,
state and measurement the discussion of interpretation problems
may be kept free from spurious assumptions.

        Park and Band$^{(12)}$ point out
that many otherwise very good texts on quantum mechanics do not give
a proper presentation of the empirical significance of the formalism.
 Following Margenau$^{(13)}$ they stress the importance of the
preparation-measurement format of experimental science. Taking this
last approach, we present here a formulation of quantum  mechanics 
that  preserves the essence of the  probabilistic  interpretation,
without introducing the Observer, who is replaced by  macroscopic 
systems. Of course the idea is not new;  see, for  example,
Heisenberg,$^{(14)}$
Weisskopf$^{(15)}$ or  Ludwig.$^{(16)}$ Our definitions of state and
observable attribute precise empirical meanings to these concepts,
independently of the mathematical formalism.
 From its definition it becomes evident that the
quantum state does indeed represent an ensemble of similarly prepared
systems ($I1$).
 As the generic state that results from such
definitions is a mixture, the traditional axioms of quantum mechanics
are not appropriate because they are expressed
in terms of state-vectors. We have therefore completed
the formulation with an alternative set of postulates in terms of density
operators. The theory that results is objective because it refers to the
objective physical reality, that is to say to the macroscopic states of
macroscopic systems.
 Within this framework the classical paradoxes of quantum mechanics are
easily resolved.

     We start the paper by giving operational definitions of state and
observable, then we postulate the mathematical formalism and its empirical 
meaning, and finally we treat some classical paradoxes of interpretation.  
 In  spite  of the lack of originality of some  of  the  ideas 
found in the formulation, we think it is worthwhile presenting a
comprehensive treatment to 
establish a  precise language for the discussion that follows.

\beginsection II. Definitions of State and Observable

     In the  definition of state presented here, as in the one given by
Jauch,$^{(17)}$ the preparation of the system determines the state.
  Here the essential role played by macroscopic systems 
within the theory is clearly exposed.
  
     The main primitive concepts that  we use are:  system,  part 
of  a system,  macroscopic observable,  macroscopic  measurement, 
classical field, interaction and time.

     A system is a {\bf macrosystem} if, associated with it, there is a 
set  of  physical quantities (macroscopic observables)  with  the  
following properties:
\item{a)} At least for limited lapses their evolution follows deterministic laws.
\item{b)} They  may  be measured simultaneously and  without  changing 
  their values.
\item{c)} Their  measurement can be performed without affecting  their 
     evolution.
  
     The fact that the macroscopic observables may be 
 measured, or   deduced  from  the  physical laws, by different 
observers independently and  without  affecting  their  evolution,
allows  us to consider the values of these macroscopic
observables as properties of the macrosystems, regardless of
the  measurements that are or are not performed.
  
     The  {\bf macrostate}  of  a system is the set of  values  of  its 
macroscopic observables.

     We  will say that the macrostates of macrosystems  represent   
the ``objective physical reality''.  Here ``objective'' means that it 
can  be empirically verified by different  observers.
  
     It is possible to state objectively that  something  happens 
only  if  some  macrostate changes.  Thus an  {\bf event}  can  be  
defined as any change of macrostate of any macrosystem.
  
     The   aim  of  any  dynamical  theory  is  to  predict   the 
probability of occurrence of a set of events $M$ with the condition 
that  another  set  of events $P$ occurs. The set of events $P$ is 
called preparation and  the set of events 
$M$ measurement. In order  to  avoid  any 
misunderstanding  it is convenient to note that ``to  predict''  is 
devoid here  of any temporal meaning.  It means that  the  theory   
yields  the probabilities independently of the actual occurrence  of the
events.   The events of $P$ and $M$ can be spread in time. Some of
the events of $P$ could be in the future of events of $M$ and vice-versa.
Nevertheless it is assumed that there is a subset of $P$, the initial
preparation, which is in the past of all events of $M$. Moreover such
initial preparation should be a sufficient one, as defined below. The
theory does assume the arrow of time.
 The  theory  does not require the existence  of  a
 physicist ``measuring''   or  ``preparing;'' the Observer is superfluous.

 We now give more precise definitions of preparation and measurement.
 
     An {\bf elemental preparation} of a system consists of:
\item{a)} An interaction of the system with a macrosystem.
\item{b)} A  set  of  conditions that must  be  satisfied
 by the macrostates of the macroscopic system during  this interaction.

     A {\bf compound preparation} of a system is made of:
\item{a)} Any number of elemental preparations of the system.
\item{b)} Any number of interactions of the system with other  systems
     previously prepared.

     {\bf Measurement}  is  any  interaction between  a  system  and  a   
macrosystem   (the  measuring  device)  which  can  change   its 
macrostate as a result of the interaction. It can happen that the
 macrosystem does not change its macrostate during the
interaction, but it is essential that the physical   
situation  allows the change to take place in order  to  consider 
the interaction as a measurement, otherwise there are no events associated
with it.
 
     The  macrostates  the  measuring device may  reach  are  the 
possible  {\bf results} of the measurement.  Usually the results  of  a 
measurement are associated to numerical values.   Any measurement 
is a preparation, but the converse is not true, because there are
preparations in which the macrosystem cannot change its macrostate.

     A  preparation is {\bf sufficient} if it determines the  probability 
distribution  of  the  results  of  any  successive  measurement. 
Intuitively  one may say that a sufficient preparation disturbs  so 
much the system that its previous history becomes irrelevant  for  
the time evolution that follows.
  
     Two  sufficient preparations are {\bf equivalent} if they  determine 
  the   same  probability  distribution  of  the  results  of   any  
measurement.    The  {\bf states}  of  a  system  are  the  classes  of   
equivalence of its sufficient preparations.

     Two   measurements  are  {\bf equivalent}  if  they   have   equal   
probability  distributions  of their results for  any  state.  The 
classes of equivalence of the measurements are the {\bf observables}.
These definitions of state and observable are similar to those given 
by Beltrametti and Casinelli.$^{(18)}$

 The state in quantum mechanics,
     as it is clear from its definition,
 is not a property of a single system.  It is a property  of 
the preparation, or, what is the same thing, of the  statistical 
ensemble  of  systems  with  equivalent  preparations. The  state   
represents  the information that is available about  the  system, 
that  is,  the conditions of the probabilities  predicted  by  the   
theory.  That  quantum-mechanical  state does  not  represent  the 
``physical  reality'' of the single system is  better  seen   
with an example:  Consider two preparations which have in  common 
the  macrosystem  but  that differ by  the  conditions  that  the 
macrostates  must fulfil ({\it e.g.} in one case some event related
to the macrosystem is assumed to occur, in the other such assumption
is not made);  assume that a system is prepared  and 
that  it satisfies both sets of conditions (which are supposed  to 
be  non-equivalent);  in such case one may assign  two  different 
states  to the same system depending on which set  of  conditions 
one  wants  to consider.  In other words the  single  system  may 
belong to different statistical ensembles. All this is very nicely
explained by Newton,$^{(19)}$ by Park$^{(20)}$ and by Tschudi.$^{(21)}$

\beginsection III. Postulates

     We can now state the postulates of the theory.  In order
not to complicate unnecessarily the exposition we give a formulation
without superselection rules. Postulates I  and  II  assure  the
existence of  states.  Postulate  III  is equivalent  to say that
once the history previous to the preparation becomes irrelevant,
nothing can be done afterwards to make it relevant again.  Postulates
IV--VIII are equivalent to the usual postulates  of  quantum mechanics
as given for  example  by  von Neumann;$^{(22)}$  postulate IV
introduces the
Hilbert space and its relation with the states; postulate V introduces the
observables; postulate VI gives the connection between the formalism and the
empirical reality; postulate VII consider compound systems and
postulate VIII
gives the time evolution of states. Finally postulate IX is a modified
version
of the reduction postulate that takes into account the fact that not all 
measurements are ideal.
 
\proclaim Postulate I. There are macrosystems.

\proclaim Postulate II. There are sufficient preparations.

\proclaim Postulate III. A preparation composed of a sufficient preparation
 followed by  another arbitrary preparation is also a sufficient one.

\proclaim Postulate IV.
     Any system has an associated Hilbert space such that
     there  is a one-to-one correspondence between the states  of  
     the system and the density operators $\rho$ with the properties:
$$	\rho = \rho ^\dagger ,\qquad
	\rho \geq 0 , \qquad {\rm Tr}\,\rho = 1. \eqno (3.1)$$

     The   mathematical  properties satisfied  by   the   density   
operators, Eq.~(3.1),  imply that they have  a  bounded  discrete 
spectrum.

     A  state is {\bf pure} if its density operator is  idempotent.  In   
this  case  the operator is a projection operator  that  projects 
into  a subspace of dimension 1.  A state is {\bf mixed} or a {\bf mixture}
if it  is  not pure.

\proclaim Postulate V.
     There  is a one-to-one correspondence between  self-adjoint   
     operators and observables. The correspondence is such that if
     $G$ is the operator corresponding to the observable $g$  and 
     if  $f$  is  a  real  function,  then  $f(G)$  is  the  operator
     corresponding to the observable $f(g)$.

\proclaim Postulate VI.
       If  a  system has been prepared in a state  with  a  density   
     operator $\rho$  the expectation value of the  measurements  of  an 
     observable $g$ with corresponding operator $G$ is given by
       $$ \langle g \rangle = {\rm Tr}(\rho G) . \eqno (3.2)$$

\proclaim Postulate VII.
     If  a  system  $S$  is  made  of  two  parts  $A$ and  $B$ with   
    corresponding  Hilbert  spaces  ${\cal H}_A$  and  ${\cal H}_B$  then
  the Hilbert space corresponding  to $S$ must  be a subspace  of
    ${\cal H}_A \otimes {\cal H}_B$.  If  an observable $g$
 is  related only to
the subsystem  $A$ and  if  its operator  of ${\cal H}_A$ is $G_A$ then its
    operator as an observable of $S$  is given by $G_S = G_A \otimes I $.

     As  a  corollary of postulates IV--VII one  gets  that  the   
density  operator $\rho_A$ of subsystem $A$ is  obtained  from $\rho$,  the 
density operator of $S$,  by contracting all indices relative to $B$,   
that is, by performing a partial trace

$$ \rho_A = {\rm Tr}^{(B)}\rho .\eqno (3.3)$$

     Consider a preparation composed of : I) the preparation of a   
state $\rho_0$,  II) no other interaction with the environment
until time $t$.  Postulate  III 
ensures  that this preparation is sufficient.  The system  will  be 
therefore  in  a state, which, because of its internal dynamics,
 is in general different  from $\rho_0$.  In 
other  words,  after  a system has been prepared in  a  state  it 
continues to have a time-dependent state $\rho(t)$.

\proclaim Postulate VIII.
     If a system is isolated or if it interacts only with  fields   
    determined by macrosystems which are insensitive to the  reaction 
    of the system,  there is a self-adjoint operator $H(t)$ (Hamiltonian 
    operator) that determines the time evolution of the  system;  the 
    time-dependent   density  operator  satisfies   the  von
 Neumann-Schr\"odinger equation
$$ i\hbar\, d\rho/dt = [H,\rho] . \eqno (3.4)$$

     We  define {\bf filter} as an observable that can assume  only   
two  values:  $0$ and $1$.  The operator of a filter is  a  projection   
operator.   The  measurement of any observable can be reduced  to 
the  measurement  of  the set of  filters  corresponding  to  its  
spectral measure.

     The  measurement  of a filter $F$ on a state $\rho$ is said  to  be
{\bf ideal} or {\bf minimal} if the density operator after the
measurement is

$$  \rho ' = F\rho F/{\rm Tr}(F\rho )  \eqno (3.5)$$
\noindent
if the result were $1$, or

$$  \rho ' = (I-F)\rho (I-F)/{\rm Tr}((I-F)\rho )   \eqno (3.6)$$
\noindent
if the result were $0$.

     When the result of an ideal measurement of $F$ is disregarded   
the new density operator is given by
$$  \rho' = F\rho F + (I-F)\rho (I-F).   \eqno   (3.7)$$
  
     Ideal measurements produce the minimum disturbance  of   
the state. The immediate repetition of an ideal measurement gives the
same result and does not modify further the state.

\proclaim Postulate IX. There are ideal measurements of any filter.
 
     Suppose  there  is  a state $\rho$ and a device  that  performs  an 
ideal measurement  of an observable which  commutes  with  the 
operator $\rho$. If the result is disregarded the interaction with the 
device  leaves  the  state unchanged.  On the other  hand  it  is 
possible  to  use the different results to  classify  the  states 
after  the  measurement.  It  is obvious that  in  this  way  the 
information  about the system is increased.  It should  be  noted 
that  a pure state cannot be reduced further by  this  procedure; 
one  may  say  that the pure states have  a  maximum  content  of 
information.  This  statement,  as  any other one of  this  paper 
referring to amount of information,   can be made quantitative if 
one   defines   the  content  of  information  of  a   state  as
   
$$I(\rho )={\rm Tr}(\rho \ln(\rho )). \eqno (3.8)$$

 It is obvious  that $I(\rho ) \leq 0$
and  that  the equality holds only for pure states.

\beginsection IV. Comments

     1) A still open problem is whether quantum mechanics applies to
macrosystems and whether the laws of macroscopic physics can be deduced as
suitable limits of the quantum-mechanical laws or if, instead, such a
program is impossible and there is a sharp boundary between microscopic and
macroscopic. The present formulation is consistent with both alternatives,
although macrosystems appear explicitly in it. To give a solution to this
problem is outside the scope of this paper; we nevertheless believe
that the first point of view is the correct one and we will assume it is so
when discussing the paradoxes, where it does make a difference.

 A substantial step toward the solution of the problem is the formulation of
R.\ Omn\`es$^{(23)}$ that does not assume the existence of
macroscopic systems from the beginning.
He first introduces the mathematical formalism, then he shows that some
systems behave classically and finally he gives an empirical meaning
to the formalism.
We consider his formulation compatible with our more traditional one.

     2) The formulation is objective because the intervention  of   
the   observer  has  been  replaced  by  the   interaction   with 
macrosystems.

     3)  Postulate  I  does not mean  that  macrosystems  are  an   
essential  part  of physical reality;  macrosystems  are  only  a 
prerequisite for acquiring any knowledge about that reality.

     4)  Our definition of state does not  distinguish   
between  pure  and  mixed states;  we define  pure state  
{\it a posteriori} by using the properties of Hilbert space.  Nor are
the empirical determinations of pure and mixed states different: 
in  both  cases  one  has  to  measure   frequency 
distributions of different observables on a statistical  ensemble 
of  equally  prepared systems.   Only after the  state  has  been 
determined does one know how pure it was. The formulation in 
terms of density operators, and not that in terms of state-vectors, is
therefore the one that follows naturally  from this definition of state. 
 This point of view is, for example, shared by Fano$^{(24)}$ and
by Park and Band.$^{(12,20,25)}$
Although it is possible to define the pure states before the introduction
of Hilbert space by using the property of convexity,$^{(26)}$ we consider
the formulation based on density operators better, because, as we will see
in Sec.~IX, there are non-pure states which are not statistical
mixtures of pure states.
 
     5) Many  formulations  of  quantum  mechanics  include   a 
reduction   (or projection)  postulate   that  implies that   any 
measurement  is ideal.  Many authors have  indicated  that  in 
general it is impossible to determine the new state prepared by a 
measurement  without considering the details of the  experimental 
arrangements; indeed, in  practice, most  measurements  are  not 
ideal ({\it e.g.}~one measures a photon by absorbing it). The weaker 
postulate  IX   only  asserts  the  theoretical  possibility  of  
performing an ideal measurement.  If a particular measurement is 
ideal  or  not  should be decided  by  analyzing  the  physical 
situation.  Margenau$^{(13)}$ advocates the rejection of the reduction   
postulate,  partially because of the previous reasons, and mainly 
because he envisages  the possibility that without the reduction
postulate the  EPR 
paradox  does not arise.  We will show that it is not so.  It  is 
maybe true that the postulate IX is superfluous, but it is 
useful and it seems to do no harm.

\beginsection V. Non-Factorizability of Quantum States

     The  state of a system composed of two parts $A$ and $B$ is  not 
determined by the states of its parts $\rho_A$ and $\rho_B$ alone.
  Only when the states of $A$ and $B$ are statistically independent the 
state of the whole system equals $\rho_A \otimes \rho_B$ . This 
property  has been called entanglement or non-separability of
quantum states.
\ d'Espagnat$^{(27)}$ uses the term separability for a different,
but related concept implying a space separation. We prefer to call it 
non-factorizability.   Actually there is nothing specially
quantum-mechanical in  this  property,  it is a  common  feature  of 
statistical  ensembles, {\it i.e.}~the marginal probability distributions
do not determine the joint probability distribution. For example a similar
situation arises also in classical statistical mechanics.  The
non-factorizability  means that the measurements of observables of $A$
are correlated to the  measurements  of observables of $B$.
 {\bf The state of  the  whole 
contains  more  information than the states  of  its  parts} (see appendix).

     It  is easy to prove that if the state of one  subsystem  is 
pure then it is statistically independent of the rest of the system.

     An  interesting  consequence of non-factorizability  is  the 
fact  that the measurement of an observable of part $A$ prepares  a 
new  state  of  part $B$.    The correlation  of  the  measurements 
yields  information  about  one  system  when  measurements   are 
performed  on  the  other one.   For  example  consider  the  ideal 
measurement of a filter $F$ of $A$, $F = F_A \otimes I$ ; the measurement
of F prepares the following state of $B$ when the result is $1$ (Eq.~3.5)
$$  \rho_B^{(1)}= {\rm Tr}^{(A)}(F\rho F) /{\rm Tr}(\rho_AF_A) \eqno (5.1)$$

\noindent
and the state

$$ \rho _B^{(0)}= {\rm Tr}^{(A)}((I-F)\rho(I-F))/{\rm Tr}(\rho_A(I-F_A))
 \eqno (5.2)$$
\noindent
when the result is $0$ (Eq.~3.6) . Note that $\rho_B$ is not modified if  
only  the  fact that there has been a measurement,  but  not  the 
result,  is  included  into the conditions that  define  the  new 
state.  The reason is that

$$ {\rm Tr}^{(A)}(F\rho (I-F)) = {\rm Tr}^{(A)}((I-F)\rho F) =
 0 \eqno (5.3)$$

\noindent
and therefore

$$ {\rm Tr}^{(A)}[F\rho F + (I-F)\rho (I-F)]=\rho_B .\eqno (5.4)$$

     If  there is statistical independence, the state of $B$ is  not   
modified even by considering the results. In that case

$$ {\rm Tr}^{(A)}(F\rho F) = {\rm Tr}^{(A)}(F_A\rho_AF_A \otimes \rho_B)
 = {\rm Tr}^{(A)}(\rho_AF_A)\rho_B \eqno (5.5)$$

\noindent
and

$$\eqalignno{
 {\rm Tr}^{(A)}[(I-F)\rho (I-F)] &= {\rm Tr}^{(A)}[(I-F_A)\rho_A(I-F_A)
\otimes \rho_B] \cr
  &= {\rm Tr}^{(A)}[\rho_A(I-F_A)] \rho_B  &(5.6)}$$

\noindent
and  therefore  it  follows  from  equations  (5.1)  and  (5.2)  that   
$\rho_B^{(1)}=\rho_B^{(0)}=\rho_B$.

\beginsection VI. The Reduction of the Wave-Packet

     Consider  a photon in a state represented by  a  wave-packet   
which  falls  upon a semireflecting mirror and  splits  into  two 
divergent  pieces.  The  photon is measured to be in one  of  the 
beams;  as  a  consequence  the wave-packet  in  the  other  beam 
disappears. Two questions arise:
  
1)   Is   this  discontinuous  change  of  the   wave-packet   in  
contradiction  with  the continuous evolution  described  by  the 
Schr\"odinger  equation  which  may also include the  measuring 
device?

2) Is not the instantaneous disappearance of a piece of the
wave-packet,   when  a  measurement  is  performed  in  another  place 
arbitrarily far away, in contradiction with relativity?

     The  origin  of  the  paradoxes is the  attempt  to  give  a   
physical reality to the quantum-mechanical state; to consider that the 
wave-function were a classical field.   The paradoxes  disappear 
once  one realizes that the statements about the  wave-packet  are   
not about the physical reality but about the predictions one  can 
make about it. The theory predicts conditional probabilities, the 
conditions   being  represented  by  the   initial   state.   The 
predictions  pertaining to the time after a measurement has  been 
performed  must  include as a condition the  occurrence  of  that 
measurement.  That  is  what  is  meant  when  one  says  that  a 
measurement  prepares a new state.  Contrary to what has been 
asserted   by   many   authors  ({\it e.g.}~von Neumann,$^{(22)}$
Wigner,$^{(28)}$ Burgos$^{(29)}$), the reduction postulate  does 
not  imply  any  kind  of different  evolution  of  the
quantum-mechanical state; it is just a prescription for  assigning 
the operator density to the new state that results when the  fact 
that  there has been an ideal measurement is included  into  the   
conditions that define the state.  The state changes because  its 
definition changes. No wonder that for von Neumann the reduction
happens in the mind of some observer; indeed to change a definition
is a mental process, but this has nothing to do with any kind of
physical evolution of the state.

 Schr\"odinger's equation rules the time evolution 
of the predictions that can be made; it is valid only between the 
preparation of the state and the measurement, because then what is being
predicted happens. Of course one can consider a bigger system that
includes the measuring apparatus; in such case the Schr\"odinger's
equation is valid until a different measurement happens. But this
is the argument of the next section.

\beginsection VII. Schr\"odinger's Cat

   What has been called ``the measurement problem'' of quantum
mechanics is dramatically illustrated by the Schr\"odinger's cat
paradox. There  is  no better presentation of this paradox  than
in  the words of Schr\"odinger himself.$^{(1,30)}$

     ``A  cat  is  placed in a steel  chamber  together  with  the 
following  hellish contraption (which must be  protected  against 
direct  interference by the cat):  in a Geiger counter there is a 
tiny amount of radioactive substance,  so tiny that maybe  within   
an  hour one of the atoms decays,  but equally probably  none  of 
them  decays.  If one decays then the counter triggers and via  a 
relay  activates  a  little hammer which breaks  a  container  of 
cyanide. If one has left this entire system for an hour, then one 
would  say that the cat is still living if no atom  has  decayed. 
The  first decay would have poisoned it.  The $\psi$-function  of  the 
entire system would express this by containing equal parts of the 
living and dead cat.

     The typical feature of these cases is that an  indeterminacy   
is  transferred from the atomic to the crude  macroscopic  level, 
which then can be decided by direct observation. \dots By itself it 
is not at all unclear or contradictory.''

     One must agree with what Schr\"odinger says.  This example  is   
not paradoxical at all unless one insists on considering that the 
quantum-mechanical  state represents the physical  reality of the
single system.  Then one  gets the absurd conclusion that the cat
is neither dead  nor alive until someone looks at it!  The truth
is instead that the death   
of the cat is completely independent on whether we look at it  or 
not.  The  cat  might be already dead,  but if one has  not  looked 
inside  the chamber one has to continue to predict a  probability
for  it being alive. It is so because the whole apparatus, including
the counter and the cat, is a macroscopic system and, therefore, always
has a determined macrostate. On the other hand quantum mechanics
only yields the probabilities of the various macrostates, and
hence {\bf it cannot predict which is the actual macrostate of
the cat}. Actually, this is the essence of the probabilistic nature
of quantum phenomena. In other words, {\bf the macrostate of a macrosystem
is not a function of its quantum-mechanical state}. What the
quantum-mechanical state determines is the distribution of macrostates.
In some cases such distribution is a sharp peak around a macrostate, in
others like in this example of Schr\"odinger the distribution is
multimodal or broad.
  
One can use the macrostate of the cat to decide (measure) whether the
radioactive substance has decayed. So two different states can
be assigned to the system depending if one imposes the  macrostate
of the cat as a condition. None of the two states is ''better''
or more ''real'' than the other one, they just correspond to different
conditions. This shows, once again, that the
quantum-mechanical state does not represent the physical reality of
the single system.

The real open question about measurement is to determine under which
conditions does a system behave as a macrosystem. A puzzling question:
How can the macrostate, which is a property of the macrosystem,
emerge from the quantum-mechanical state, which is not?

\beginsection VIII. Quantum Mechanics and Determinism

     It  has  been  stated  many  times  (see  for  example   von 
Neumann$^{(22)}$) that the time evolution of a system is deterministic 
except during measurements, because the evolution of the state is 
governed  by  a  differential equation.   We  have  also  here  a 
confusion  between the evolution of the {\bf system} and the  evolution 
of the {\bf predictions} (state)  that one can make about it. The
question of whether 
the  evolution  of a non-macroscopic system is deterministic  or  not, 
cannot  even  be posed because in this case there are  no  events   
between the preparation and the measurement,  while the evolution 
of a  system  is a succession of events.   If the  system  has  a 
macroscopic  part  it is clear that in general the  evolution  of 
events  is not deterministic (think of  Schr\"odinger's  cat).  The 
probabilities of events are completely determined by the  initial 
state, not so the events that actually happen.

\beginsection IX. EPR Paradox I: Are Density Operators Superfluous?

     The  famous  paper by Einstein,  Podolsky  and  Rosen$^{(3)}$ is 
  without  doubt  one of the most interesting and  discussed  works 
about the interpretation of quantum mechanics.  The problem  they 
have  proposed has become known as the EPR paradox, in spite of the fact
that in no place of their paper the  authors   
claim  to  have found a paradox.  On the  other  hand, several quite 
different paradoxes have been based on the same physical situation
presented in this piece of work.  As  a 
result  there is always an ambiguity when someone refers  to  the 
``well  known  EPR paradox''.  The core of EPR's  argument  is  the 
following  example,  presented here in a simplified form  due  to 
Bohm.$^{(31)}$
  
     Consider  a  system composed of two particles of  spin $1/2$. 
Assume that the Hamiltonian commutes with the total spin
$\vec S=\vec S_A + \vec S_B$. The system is prepared initially in
 a  metastable  bound 
state   of   total  spin  $S=0$.   After  some  time   the   system 
disintegrates, the two particles separate and no longer interact, 
but the total spin maintains its initial value $S=0$.  The state of   
both  spins is pure, thus it can be factorized from the  space  part 
and it is represented by the state vector

$$ |\psi\rangle ={1\over\sqrt 2}\bigl(|\hbox{${1\over2},-{1\over2}$}
\rangle -|\hbox{$-{1\over2} ,{1\over2}$}\rangle\bigr). \eqno (9.1)$$

     The  density matrix of each spin is obtained by the  partial   
trace:

$$\rho_A = {\rm Tr}^{(B)}|\psi\rangle\langle\psi |= {1\over2} I;
\qquad\rho_B = {\rm Tr}^{(A)}|\psi\rangle\langle\psi |= {1\over2} I. 
\eqno (9.2)$$

     So both spins are completely unpolarized.  There is, though, a   
complete  correlation between the measurements of the  components 
of  $\vec S_A$  and $\vec S_B$ in the same arbitrary direction 
 $\hat n$.  In  fact  the 
results  are  always opposite because  ( $\vec S_A\!\cdot\hat n +
\vec S_B\!\cdot \hat n )|\psi\rangle = 0$ . Then by measuring
$\vec S_A\!\cdot \hat n$ one prepares the spin $B$ in the pure  state 
with eigenvalue of $\vec S_B\!\cdot\hat n$ opposite to the result of the
 measurement of  spin $A$.
  
     Many  versions  of the EPR paradox are based  on  the  wrong 
assumption, sometimes implicit, that the use of density operators in the
formalism of quantum mechanics  
is  superfluous$^{(29,32)}$.  More precisely the  following  statement, 
that has been called microrealism,$^{(5)}$ is assumed to be true:

{\rightskip=\parindent
\item{($F$)} ``In a mixed state the density operator describes an ensemble 
     of  systems  prepared with the same  procedure,  but  it  is 
     always  possible to assume that each particular  element  of 
     the ensemble is in its own pure state.''

}
     By  the way, the belief that $F$ was true is the origin  of  the   
name ``mixture'' given to the non-pure states. Actually EPR did not 
make this wrong assumption.  The proof of the falsity of $F$ is  as   
follows:

     Assume that the statement $F$ were true.  Then the spin $B$ will   
be in a pure state $r$.  Now,  by measuring $\vec S_A\!\cdot \hat n$,
an action that in no way  can  modify  the {\bf pure} state $r$
(see sec.~V), one prepares spin $B$ in  an 
eigenstate  of $\vec S_B\!\cdot\hat n$.  It follows that $r$ is such
eigenstate.  But since  $\hat n$ is arbitrary, $r$ must be eigenstate of
any component of 
$\vec S_B$, in contradiction with the commutation relations.

     Different authors$^{(33,34)}$ analyzing this same  example,  also 
conclude that there are situations that cannot be described  with 
wave-functions but only with density matrices. \ d'Espagnat$^{(27)}$   
finds the same thing and calls these cases improper mixtures  (or 
mixtures  of the second kind),  but insists in  considering  that 
only  pure states are ``true'' quantum states; this very common point
of view has the unpleasant consequence that there would be 
 systems that have no state or, alternatively that very well  defined 
pieces of physical reality (as particle $B$) could not be called systems.

\beginsection X. EPR Paradox II: Is Quantum Mechanics Complete?

     The  aim  of EPR in their original paper was to  prove  that 
``the quantum-mechanical description of physical reality given  by 
wave  functions  is not complete''.   By complete EPR  meant  that 
``every element of the physical reality must have a counterpart in 
the physical theory.'' Apart from the argumentation that EPR gave
in their paper it is clear that what they wanted to prove is {\bf true}.
We have seen in Sec.~VII that in general the quantum state of a
macroscopic system, even if it is pure, does not determine its
macrostate, but only a probability distribution of macrostates.
As macrostates compose the objective physical reality the description
given by wave-functions (or density operators) does not agree
with EPR's definition of complete. Now return to the EPR's argumentation.
They start by giving the following criterion  of reality:

{\rightskip=\parindent
\item{($R$)} ``If,  without  in any way disturbing a system,  we  can  
predict with certainty ({\it i.e.}~with probability equal to unity) the   
values  of a physical quantity, then there exists an  element  of 
physical  reality corresponding to this physical  quantity''.

}
  Then they  show  that  the 
impossibility  of predicting simultaneously the values of  observables
 that  do  not  commute implies  that  the  following  two   
statements cannot be both false:

{\rightskip=\parindent
\item{(1)}  The quantum-mechanical description of reality given by wave-functions is not complete.
\item{(2)}  When the operators corresponding to two physical  quantities 
     do  not commute the two quantities cannot have  simultaneous 
     reality.

}  
     EPR give then an example from which they pretend to conclude  
that statement (2) is false ({\it i.e.}~non commuting  observables  may 
have simultaneous physical reality),  and that therefore (1) must   
be  true  (QM is not complete).   EPR's argumentation is  composed  of 
three parts.  First they ascertain that it is possible to prepare 
a  pure  state  of  system $B$ by  measuring  system  $A$.  In  the   
simplified  version  introduced in section IX this  part  of  the 
argument will read as:

{\rightskip=\parindent  
\item{($A1$)}      Suppose  that  $\vec S_A\!\cdot\hat a$ is measured and
that the result is $\alpha$ ($\alpha=\pm 1/2$). Then the system $B$ is left
in  the spin state $|\vec S_B\!\cdot\hat a=-\alpha\rangle$; take instead 
another direction $\hat b$ and let   
the result be $\beta$, then the system $B$ is left in the spin state 
    $|\vec S_B\!\cdot\hat b=-\beta\rangle$.

}
\vskip 12pt
     Here comes the central point in their  argument.  They say:

{\rightskip=\parindent  
\item{($A2$)} ``\dots On the other hand,  since at the time of  measurement 
     the two systems no longer interact,  no real change can take 
     place  in the second system in consequence of anything  that  
     may be done to the first system. This is of course, merely a   
     statement  of  what is meant by the absence  of  interaction 
     between the two systems.''

\item{($A3$)} ``Thus it is possible to assign two different wave functions 
     (\dots) to the same physical reality ( the second system after 
     the interaction with the first)''.  

}
 
Then they use the reality criterion to conclude that both
$\vec S_B\!\cdot\hat a$ and $\vec S_B\!\cdot\hat b$ simultaneously have
physical reality, and therefore
as  $\vec S_B\!\cdot\hat a$  does not commute with $\vec S_B\!\cdot\hat b$
statement  (2)  is contradicted and (1) must be true.
Up to here EPR's argument.  But there is more;  given the arbitrariness
of $\hat b$, with the same reasoning one gets that
any component of $\vec S_A$ and $\vec S_B$
must have  physical reality and therefore must be predetermined
since the  time  when the two systems were interacting. Here  an  
unavoidable difficulty appears: since this last statement implies 
that  Bell's inequalities are satisfied,  it contradicts  quantum  
mechanics.$^{(35,36,7)}$ As different authors$^{(37,38,39)}$ claim that
the   proof of Bell's inequalities is based on the probability theory  of 
Kolmogorov that perhaps is not valid in this case, I present here   
a proof of one of Bell's inequalities that does
not make use of probabilities. Consider 
a number $N$ of pairs of spins prepared as above. Define $A(\hat a,i)$ and
$B(\hat b,i)$  as  the predetermined results of $2\vec S_A\!\cdot\hat a$
and $2\vec S_B\!\cdot\hat b$  of the pair labelled by i. Define also 
$P(\hat a,\hat b)$ as
  
$$ P(\hat a,\hat b) = 1/N\sum_i A(\hat a,i)B(\hat b,i). \eqno (10.1)$$

     Given  that $|A|=|B|=1$ and that $B(\hat b,i)=-A(\hat b,i)$ it
immediately  follows that
  
$$ |P(\hat a,\hat b)-P(\hat a,\hat c)|\leq 1+P(\hat b,\hat c).
 \eqno (10.2)$$
  
     But, if quantum predictions are of any use, as $N$ goes to $\infty$
the quantity $P(\hat a,\hat b)$ must approach $\langle 2\vec S_A\!
\cdot\hat a \ 2 \vec S_B\!\cdot\hat b\rangle = -\hat a\cdot\hat b$  and
this last expression does not satisfy the inequality (10.2).  Therefore
there is something wrong in the argument of EPR because it  contradicts 
quantum mechanics;   so EPR did not prove that quantum  mechanics 
was incomplete. After all the EPR argument was indeed a paradox.

     This last argument reveals a main difference between classical
probabilistic theories and quantum mechanics. In classical probability for
any two random variables a joint probability distribution exists. In quantum
mechanics it is not so. Only {\bf compatible} observables may have a joint
probability distribution.$^{(36)}$

\beginsection XI. EPR Paradox III: What Went Wrong?
  
     The  interesting question now is:  what is wrong in the 
argument of EPR? They themselves state the possibility that that their
reality criterion were not valid, and indeed the problems have been usually
attributed to such criterion.$^{(40,41)}$ It is our opinion that there is
nothing wrong with EPR's reality criterion. In fact the statement $A3$
of EPR's argumentation, which appears before the criterion is used, already
contains an absurdity: what sense may have to assign two different
{\bf predictions}
to the same physical reality? Since the statement $A1$ is just a consequence
of standard quantum mechanical rules the troubles must be attributed to
$A2$. If $A2$ were right the only alternative left would be to assume that
the hypothesis of isolated systems cannot be made! But this has immense
consequences as relativity implies that two systems can be isolated from
each other during some time interval by separating them in space (locality).

     For  d'Espagnat$^{(27,42)}$ the non-factorizability  of  quantum 
states indeed implies non-locality and supraluminal influences.

Let us analyze $A2$ carefully. Of course $A2$ implies the isolation of
the systems, but the converse is not true. $A2$ is too strong. What
is really needed by the absence of interaction between two systems
is:

{\rightskip=\parindent
\item{($A2^\prime$)} Nothing that may be done to the first
system can produce a change in the second system.

}

\noindent
And indeed no measurement performed on system $A$ can change the
state $\rho_B={1\over2}I$ of spin $B$, see Eq.~(5.4).
 
 But then, what is the
 meaning of the reduction of the state $\rho_B$ to the states
\hbox{$|\vec S_B\!\cdot\hat a=-\alpha\rangle$}?  This  change   
is  not due to the interaction of spin $A$ with the  measuring 
device  but to the fact that the result of such a measurement  has 
been  {\bf included} into the {\bf definition} of the state of $B$.
  Therefore  with  this  new  condition  the  statistical ensemble which
is represented by the state of spin $B$ is different from the original 
one  (half of the spins have been excluded from it!),  as  it  is 
different   from  the  ensemble  which  is  obtained   when   the 
(incompatible)  condition  that  specifies  the  result  of   the 
measurement of spin $A$ in another direction is included.
It is therefore wrong to claim that those three situations are the 
``same  physical reality''.  So $A3$ does not follow from $A1$
and $A2^\prime$.
There is no contradiction  with locality.  It is the reduction of
the  wave-packet paradox again.
  
    In  the  EPR  paper  the  paradox  arises  because  it   was  
overlooked that to specify a result of the measurement of system $A$  
implies the inclusion of a new condition into the definition of the 
state of system $B$; that, of course, ``disturbs'' the system ({\it
i.e.}\ changes its state) and hence the reality criterion cannot be applied.

	In summary, if the measurement of system $A$ is performed
and the result is disregarded then the state of system $B$ does not change
(in accord with $A2^\prime$), but the reality criterion cannot be applied
because there is no prediction with certainty. In order to have probability
equal to unity the result of the measurement of $A$ has to be included as a
condition, but then, as the state of $B$ changes, the reality criterion
cannot be applied to the original physical situation. Of course it can be
applied to the new state that results. That is to say, a component of
spin $B$ acquires physical reality only if the measurement of the same
component of spin $A$ is actually performed and the result is taken
into account. Therefore it is impossible to simultaneously give
physical reality to a different component of spin $B$, because the
corresponding two components of spin $A$ are not compatible.

  To claim that non-factorizability implies supraluminal influences
is  to  confuse  a logical fact  (the  inclusion  of  a 
condition  into  the  definition  of a  state)  with  a  physical 
interaction:  the hypothetical influence of measurement of system $A$ 
on system $B$.

     A three particle extension of the EPR problem, due to
Greenberger, Horne and Zeilinger has shown more directly that to apply
the reality criterion the way EPR did, contradicts quantum
mechanics.$^{(43,41)}$
This new case can be also analyzed with the arguments given above.

	What did lead to the mistake? In $A1$ EPR tell us that the state
of $B$ changes, in $A2$ they say that such change is not {\it real}, that is
they are implicitly assuming that there is some underlying reality which
is not described by the quantum state, {\it i.e.}~that there is some kind of
different, objective state, the ``real one'',
representing the physical situation of system $B$; of course, such an
objective state could not be modified by the measurement of system $A$. The
EPR paradox shows that similar assumptions contradict quantum mechanics
because they lead to Bell's inequalities. In this sense quantum mechanics
cannot be completed.

\beginsection XII. EPR Paradox IV: Some Replies
 
        In the light of the previous discussion we will try to
analize Bohr's reply to the EPR paper which appeared shortly
after.$^{(40)}$ Few words about Bohr's philosophy before
considering the central  point of the paper. As Bohr was very
well aware of the ambiguous empirical content that the
wave-function had in the traditional formulation, he always followed
the recommendation of  analyzing the whole experimental arrangement
when discussing the paradoxes of interpretation. But what surely
is a good measure of caution was elevated by Bohr to the category
of a postulate, denying even the possibility of giving a precise
empirical meaning to quantum states. For example, he writes in the
cited reply: ``In accordance with this situation there can be no
question of any unambiguous interpretation of the symbols of quantum
mechanics other than that embodied in the well-known rules which
allow to predict the results to be obtained by a given experimental
arrangement described in a totally classical way,\dots ''. Our
formulation of quantum mechanics shows that this extreme
instrumentalist point of view was unjustified.
 
        Most of Bohr's reply is devoted to obtain the following
result that is given here in a form appropriate for Bohm's two
spins example. By measuring a component of spin $A$ the same component
of spin $B$ can be predicted with certainty, but an orthogonal component
of spin $B$ is completely indeterminated. If one chooses to measure an
orthogonal component of spin $A$ then the same component of spin $B$ can
now be predicted, but the previous component becomes indeterminate.
Afterwards follows the central point of Bohr's argument. We prefer
to reproduce it literally as it has been interpreted in different
ways.$^{(5,7,44)}$ Bohr writes:

``From our point of view we now see that the wording of the
above-mentioned criterion of physical reality proposed by Einstein,
Podolsky and Rosen contains an ambiguity as regards the meaning of
the expression ``without in any way disturbing a system.'' Of course
there is in a case like that just considered no question of a mechanical
disturbance of the system under investigation during the last critical
stage of the measuring procedure. But even at this stage there is
essentially the question of {\it an influence on the very conditions
which define the possible types of predictions regarding the future
behavior of the system.} Since these conditions constitute an 
inherent element of the description of any phenomenon to which the
term ``physical reality'' can be properly attached, we see that the
argumentation of the mentioned authors does not justify their conclusion
that quantum-mechanical description is essentially incomplete.''

        Our interpretation of Bohr's argument is the following. Bohr
states that the conditions that define the possible types of predictions
that can be made are part of the ``physical reality'' of system $B$,
which therefore is different when incompatible observables of system
$A$ are measured. So the argumentation of EPR does not go through.
Within our interpretation of quantum mechanics 
the above argument would be incorrect because of a subtle but essential
point. What is determined by the ``physical reality'' of system $B$
is not the {\it type} of predictions that {\it could} be made, but
the predictions that {\it can} be made. In the first case the {\it type}
of measurement that is performed on system $A$ ({\it i.e.}~what
kind of measuring apparatus is acting upon $A$) would be part of the
``physical reality'' of system $B$. If that were the case it would be very
hard to sustain that there is no mechanical connection between systems
$A$ and $B$, as we would have an explicit contradiction of $A2^\prime$.
 Instead in our interpretation what makes the change
 of ``physical reality'' (state) of 
system $B$ is not the physical process of measurement on  system $A$,
but to include as a condition the result of such measurement. Bohr's
argumentation could be made valid within our interpretation giving 
a different wording to the phrase in italics:``{\it \dots conditions which
define the predictions \dots}'', but still an explanation of why the change
of ``physical reality'' of system $B$ is not a ``mechanical
disturbance'' would be missing.

        A final comment about Bohr's reply; he explicitly rejects the
criterion of physical reality of EPR, but what his argument actually shows
is that the criterion cannot be applied because system $B$ has been 
``disturbed'' by the measurement on system $A$.

 We see that what is
very difficult to explain with the traditional formulation becomes
simple and clear with ours.

 Following Margenau,$^{(13)}$ de Muynck$^{(5)}$ states that an   
instrumentalist  interpretation  of quantum  mechanics  which  is 
compatible  with locality is possible,  provided  the  projection 
postulate is rejected,  because, they claim, then the EPR paradox 
cannot be formulated.  We have two comments:  1) It is an illusion 
to  believe  that by rejecting the projection postulate  the  EPR 
problem cannot be posed.  In the Bohm version, the very existence 
of the perfect correlation,  precisely because both particles are 
not interacting,  implies that any measurement of spin $A$, ideal
or  not,  provided  the measuring device does not  interact  with   
spin  $B$,  prepares  spin $B$ in a pure state if the result  of  the   
measurement of spin $A$ is specified. This is completely independent 
of  the projection postulate.  2) There is no need to reject  the 
weak version of the projection postulate, because the   
origin  of  the problems is not in the postulate itself  but  in  its 
objectivist interpretation; that is to assume that the collapse is
something that happens at the physical level and not at the logical
level as it is in our interpretation.
  
     Considering  the  EPR  situation  Costa  de   Beauregard$^{(45)}$
correctly  points  out that the statement ``the first of the  two 
measurements  instantaneously  collapses the other  substate'' is 
unacceptable because it is manifestly noncovariant.  He  proposes   
to  eliminate from the theory the intermediate concept  of  state 
and to use only transition amplitudes.  Actually,  once again, it  
is  not  the  concept  of state in  itself  but  its  objectivist 
interpretation the origin of the problems. The ``reduction'' of the  
state of $B$ ({\it i.e.}~the inclusion of the result of a measurement of 
$A$ in the definition of the state of $B$) is an {\it a}-temporal  fact. It
is  perfectly valid to consider the state of $B$ ``collapsed''  even 
before  the  measurement of $A$ has taken place,  provided we impose the  
condition that this measurement will eventually be performed and its result
taken into account. Which measurement does ``collapse'' the other substate
is arbitrary: it depends on which one of the two measurements  one 
wants to consider as a condition for the probability of the other one.

\beginsection XIII. EPR Paradox V: Does Non-Factorizability Imply
Non-Locality?
 
        The state in quantum mechanics represents the preparation,
{\it i.e.}~the information that is available about the system. As
the state determines the probabilities of the measurements the information
should be considered encoded into the system itself. Now imagine
a system with state $\rho$ composed by two parts $A$ and $B$, with
states $\rho_A$ and $\rho_B$ respectively. There is no problem
in assuming that the information represented by $\rho_A$ is encoded
in subsystem $A$, while the information of $\rho_B$ is encoded in part
$B$, but where is the information pertaining to $\rho$ that is not
included in $\rho_A\otimes\rho_B$ encoded? {\bf In both parts taken
together}.
Therefore the complete system encodes more information than its parts
(see appendix).

  By analyzing the EPR situation d'Espagnat$^{(46)}$ obtains that quantum
mechanics is incompatible with the simultaneous assumption of realism,
inductive inference and Einstein locality. From the assumption of realism
and the existence of a perfect correlation, even when the systems are
isolated, he concludes that some property must be already present in each
subsystem A and B before the measurement were performed. Using inductive
inference the presence of the property can be extrapolated to those systems
that have not been measured. That is all what is needed in order to derive
the validity of one of Bell's inequalities, which contradicts the
predictions of quantum mechanics. Let us analyze carefully the previous
argument. In order to deduce the existence of a preexistent property in
each subsystem two hypothesis are needed: a) matter has encoded
into itself the information that together with the physical laws determines
its behavior (realism), b) such information is encoded locally. Alone,
the first hypothesis allows to deduce the existence of a property in
the composed system (the correlation), but not in each part separately.
 As discussed in the beginning of this
section the second hypothesis seems to be in direct contradiction with the
non-factorizability of quantum states, therefore it should be dropped
instead of such fundamental assumptions as realism, inductive inference or
Einstein locality.

The EPR and GHZ$^{(43)}$ paradoxes show that quantum mechanics is
incompatible with the assumption of locally encoded information. In
this sense we may say that non-factorizability of quantum states
does imply some kind of non-locality.

    On the other hand quantum  mechanics 
is consistent with  the locality of {\bf events}.  
 Consider  two events $a$  and  $b$  happening 
respectively  in  two systems $A$ and $B$  which contain  macroscopic   
parts.  In order to verify that event $a$ is cause of event $b$ (or $b$
effect of $a$) one has to:  (1) prepare both systems  independently   
and  preserve the independence of states up to the occurrence  of 
event  $a$,  (2)  verify  that the occurrences of  the  events  are 
correlated.   The statistical independence is essential otherwise  
the  correlation  could be attributed to a common cause  of  both 
events. 

     How do these last concepts apply to the example of EPR?  The 
correlation between measurements of $\vec S_A\!\cdot\hat n$ and
$\vec S_B\!\cdot\hat n$ does not  imply   
a  causal  connection because the preparations of the  states  of 
spins $A$ and $B$ are not independent. Moreover, if the same states of 
spins  $A$  and  $B$ were prepared independently there  would  be  no 
correlation between the measurements.  Therefore the  correlation 
should  be attributed to the common preparation of both  systems.   
This  is  also  revealed  by  the  impossibility  of  using   the 
correlation  between measurements to send messages.  In fact  the   
probabilities  of  all  measurements on  spin  $B$  are  completely 
determined by its state $\rho_B$  which is not affected at all by  the 
measurements performed on spin $A$ when their results are  ignored. 
In other words it is impossible from the sole observation of spin 
$B$ to infer if the spin A has been measured or not.  In conclusion, 
there  is  no  contradiction  between  the  correlation  and  the
isolation of the spins.

\beginsection XIV. Conclusion
  
     The three paradoxes that were analyzed in this paper have  a 
common  origin:  the attribution to quantum-mechanical states  of   
properties  that  they  do  not  actually  have, such as being  a 
representation of the physical reality, 
 being a property of the single
system or being an ensemble of hidden realistic states.
  Our definition of state keeps that  concept 
free  from superfluous assumptions and at the same time  allows a   
probabilistic  interpretation of quantum mechanics that  is  also 
objective. We show that quantum mechanics,
notwithstanding it cannot be completed with realistic states,
is not complete in the sense
of EPR, because macroscopic observables always have determined values
and  quantum mechanics only yields probabilities for such values.
 We also show that  quantum-mechanical correlations are not in conflict
with realism or with Einstein locality, but that they imply non-local
encoding of information.

\beginsection Acknowledgements

     I  have  to thank Drs.\ A.~K\'alnay, R.~Tello, V.~Yartsev, C.~Mendoza and
M.~Garc\'\i\-a Sucre for  many discussions,  for  reading  the manuscript
and  for their  sharp criticism.

\vfill\eject

\beginsection Appendix

In this appendix we prove some results about the content of information
of composed systems. Consider a system composed of two parts $A$ and
$B$ with states $\rho_A$ and $\rho_B$. First we show that if the two
systems are statistically independent then the content of information
is additive.
$$\eqalignno{I(\rho_A \otimes \rho_B) &= {\rm Tr}
	[\ln(\rho_A\otimes\rho_B)\rho_A\otimes\rho_B]\cr
	&={\rm Tr}[(\ln\rho_A\otimes I + 
	I\otimes\ln\rho_B)\rho_A\otimes\rho_B]\cr
	&={\rm Tr}(\ln\rho_A\rho_A\otimes\rho_B) +
	  {\rm Tr}(\rho_A\otimes\ln\rho_B\rho_B)\cr
	&= {\rm Tr}^{(A)}(\ln\rho_A \rho_A) +
	  {\rm Tr}^{(B)}(\ln\rho_B \rho_B)\cr
	&= I(\rho_A) + I(\rho_B)  &(A.1)}$$

Now we will show that the state of the composed system that has the
minimum content of information is $\rho_A\otimes\rho_B$.
Let
$$ \rho_A = \sum_i p^{A}_i |u_i\rangle \langle u_i| \eqno (A.2)$$
$$ \rho_B = \sum_i p^{B}_i |w_i\rangle \langle w_i|. \eqno (A.3)$$

A generic state of the composed system is
$$ \rho = \sum_i p_{\alpha} |e_{\alpha}\rangle \langle e_{\alpha}|
 \eqno (A.4)$$
where
$$ e_{\alpha} = \sum_{ij} U_{\alpha,ij} u_i\otimes w_j \eqno (A.5) $$
and $U_{\alpha,ij}$ is some unitary matrix.
$$ \sum_{\alpha} U_{\alpha,ij} U^*_{\alpha,kl} = \delta_{ik}\delta_{jl}
  \eqno (A.6) $$
The conditions ${\rm Tr}^{(B)}\rho = \rho_A$ and 
${\rm Tr}^{(A)}\rho = \rho_B$ become
$$ \sum_{\alpha j} p_{\alpha} U_{\alpha,ij}U^*_{\alpha,kj} = \delta_{ik}
  p^A_i \eqno (A.7) $$
and
 $$ \sum_{\alpha i} p_{\alpha} U_{\alpha,ij}U^*_{\alpha,il} = \delta_{jl}
  p^B_j \eqno (A.8) $$

In order to obtain the state of minimum content of information one has
to minimize $I(\rho)=\sum_{\alpha}p_{\alpha}\ln p_{\alpha}$ with the
conditions (A.6), (A.7) and (A.8). Using the method of Lagrange one has
to minimize the quantity $S$
$$ \eqalignno {S = &\sum_{\alpha}p_{\alpha}\ln p_{\alpha}\cr &+
	\sum_{ik} a_{ik}(\delta_{ik} p^A_i - \sum_{\alpha j}p_{\alpha}
	U_{\alpha,ij}U^*_{\alpha,kj}) \cr &+ \sum_{jl} b_{jl}( \delta_{jl}
	p^B_j - \sum_{\alpha i}p_{\alpha}U_{\alpha,ij}U^*_{\alpha,il})\cr &+
	\sum_{ijkl} c_{ijkl}(\delta_{ik}\delta_{jl}-\sum_{\alpha}
	U_{\alpha,ij}U^*_{\alpha,kl}).  &(A.9)}$$
This yields the equations
$$0= 1 + \ln p_{\alpha} - \sum_{ijk} a_{ik} U_{\alpha,ij}U^*_{\alpha,kj}
	-\sum_{ijl} b_{jl} U_{\alpha,ij}U^*_{\alpha,il}, \eqno (A.10)$$
$$0= -\sum_k a_{ik}p_{\alpha}U^*_{\alpha,kj} -\sum_l b_{jl}p_{\alpha}
  U^*_{\alpha,il} - \sum_{kl} c_{ijkl} U^*_{\alpha,kl} \eqno (A.11)$$
and
$$0= -\sum_i a_{ik}p_{\alpha}U_{\alpha,il} -\sum_j b_{jl}p_{\alpha}
  U_{\alpha,kj} - \sum_{ij} c_{ijkl} U_{\alpha,ij}. \eqno (A.12)$$

Now one has only to verify that $\rho=\rho_A\otimes\rho_B$ fulfils
equations (A.6), (A.7), (A.8), (A.10), (A.11) and (A.12). The state
$\rho_A\otimes\rho_B$ is obtained setting $\alpha=(m,n)$, 
$p_{mn}= p^A_m p^B_n$ and $U_{mn,ij}=\delta_{mi}\delta_{nj}$. Equations
(A.6), (A.7) and (A.8) are obviously satisfied. The other three equations
are also fulfilled if one sets
$$ a_{mn}= (\ln p^A_m +1/2)\delta_{mn}, \eqno (A.13)$$
$$ b_{mn}= (\ln p^B_m+ 1/2)\delta_{mn} \eqno (A.14)$$
and
$$ c_{ijkl}= -p^A_i p^B_j(\ln p^A_i + \ln p^B_j + 1)\delta_{ik}\delta_{jl}.
	\eqno (A.15)$$

Therefore for any state $\rho$ of the composed system
$$ I(\rho)\ge I(\rho_A) + I(\rho_B).  \eqno (A.16)$$

\vfill\eject

\beginsection References

\item{1.}   E.~Schr\"odinger, {\it Naturwissenschanften} {\bf 23},
807 (1935).

\item{2.}   W.~Heinsenberg, {\it The Physical Principles of Quantum 
Mechanics} (Dover Publications, Chicago, 1930), p.~39.

\item{3.}   A.~Einstein,  B.~Podolsky and N.~Rosen, {\it Phys.\ Rev.}\ {\bf
47}, 777 (1935).

\item{4.}   F.~Selleri,  G.~Tarozzi, {\it Rivista del Nuovo Cimento}
 {\bf 4}(2), 1 (1981).
  
\item{5.} W.~M.~de Muynck, {\it Found.\ Phys.}\ {\bf 16}, 973 (1986).

\item{6.} J.~G.~Cramer,  {\it Rev.\ Mod.\ Phys.}\ {\bf 58}, 647 (1986).

\item{7.} H.~P.~Stapp, {\it Found.\ Phys.}\ {\bf 21}, 1 (1991).

\item{8.} D.~Home and M.~A.~B.~Whitaker, {\it Phys.\ Rep.}\ {\bf 210},
     223 (1992).

\item{9.} L.~E.~Ballentine, {\it Rev.\ Mod.\ Phys.}\ {\bf 42}, 358 (1970).

\item{10.} N.~D.~Mermin, {\it Rev.\ Mod.\ Phys.}\ {\bf 65}, 803 (1993).

\item{11.} M.~Bunge, in {\it Quantum Theory and Reality},
       M.~Bunge ed.~(Springer-Verlag, Berlin, 1967), p.~105.

\item{12.} J.~L.~Park and W.~Band, {\it Found.\ Phys.}\ {\bf 22},
 657 (1992).

\item{13.} H.~Margenau, {\it Phys.\ Rev.}\ {\bf 49}, 240 (1936).

\item{14.} W.~Heisenberg, {\it Physics  and Philosophy} (George  
     Allen \& Unwin LTD, London, 1958), pp.~46, 114.
   
\item{15.} V.~Weisskopf,  letter to the editor in {\it Sci.\ Am.}
\ {\bf 242}(5), 8 (1980).
  
\item{16.} G.~Ludwig, {\it Foundations  of  Quantum  Mechanics}
     (Springer-Verlag,  Berlin, 1982).

\item{17.}  J.~M.~Jauch,  {\it Foundations  of  Quantum  Mechanics}
        (Addinson-Wesley, Reading Massachusetts, 1968) p.~60.

\item{18.} E.~Beltrametti and G.~Casinelli, {\it The Logic of Quantum
         Mechanics} in ``Encyclopedia of Mathematics,'' Vol.~15, 
         (Addison-Wesley, Reading Massachusetts, 1981) pp.~137--141.

\item{19.} R.~G.~Newton, {\it Am.\ J.\ Phys.}\ {\bf 48}, 1029 (1980).

\item{20.} J.~L.~Park, {\it Am.\ J.\ Phys.}\ {\bf 36}, 211 (1968).

\item{21.} H.~R.~Tschudi, {\it Helv.\ Phys.\ Acta} {\bf 60}, 363 (1987).

\item{22.} J.~von Neumann, {\it Mathematische Grundlagen der
 Quantenmechanik} (Springer-Verlag, Berlin, 1968).

\item{23.} R.~Omn\`es , {\it Ann.\ Phys.} (N.Y.) {\bf 201}, 354 (1990).

\item{24.} U.~Fano, {\it Rev.\ Mod.\ Phys.}\ {\bf 29}, 74 (1957).

\item{25.} W.~Band and J.~L.~Park, {\it Found.\ Phys.}\ {\bf 1}, 133 (1970).

\item{26.} Ref.~18 p.~139.

\item{27.} B.~d'Espagnat, {\it Phys.\ Rep.}\ {\bf 110}, 201 (1984).

\item{28.} E.~P.~Wigner, {\it Am.\ J.\ Phys.}\ {\bf 31}, 6 (1963).

\item{29.} M.~E.~Burgos, {\it Found.\ Phys.}\ {\bf 14}, 739 (1984).
 
\item{30.} Translation given by Jauch in Ref.~17, p.~185.

\item{31.} D.~Bohm, {\it Quantum Theory} (Prentice-Hall, New York,
 1951), p.~614.

\item{32.} M.~A.~B.~Whitaker and I.~Singh, {\it Phys.\ Lett.}\ {\bf 87A},
 9 (1981).
  
\item{33.} C.~Cantrell and M.~Scully, {\it Phys.\ Rep.}\ {\bf 43},
 499 (1978).

\item{34.} Ref.~18 pp.~11, 71.

\item{35.} J.~S.~Bell, {\it Physics} {\bf 1}, 195 (1964).

\item{36.} A.~Fine, {\it  Phys.\ Rev.\ Lett.}\ {\bf 48}, 291 (1982).

\item{37.} I.~Pitowsky, {\it Phys.\ Rev.\ Lett.}\ {\bf 48}, 1299 (1982).

\item{38.} S.~P.~Gudder, {\it Found.\ Phys.}\ {\bf 14}, 997 (1984).

\item{39.} N.~S.~Todorov, {\it Ann.\ Fond.\ L.\ de Broglie}, {\bf 10},
 49(1985).

\item{40.} N.~Bohr, {\it Phys.\ Rev.}\ {\bf 48}, 696 (1935).

\item{41.} N.~D.~Mermin, {\it Am.\ J.\ Phys.}\ {\bf 58}, 731 (1990).

\item{42.} B.~d'Espagnat, {\it Found.\ Phys.}\ {\bf 11}, 205 (1981).

\item{43.} D.~M.~Greenberger, M.~Horne, and A.~Zeilinger, 
       in {\it Bell's Theorem, Quantum Theory, and Conceptions of the
       Universe} M.~Kafatos ed. (Kluwer Academic, Dordrecht, 1989),p.~69.

\item{44.} D.~Bohm and Y.~Aharonov, {\it Phys.\ Rev.}\ {\bf 108}, 1070
 (1957).

\item{45.} O.~Costa de Beauregard, 
     {\it Lett. Nuovo Cimento} {\bf 36}, 39 (1983).

\item{46.} B.~d'Espagnat, {\it Sci.\ Am.}\ {\bf 241}(5), 128 (1979).

\bye